\documentstyle[11pt,psfig]{article} 
\newcommand{\beq}{\begin{equation}} 
\newcommand{\eeq}{\end{equation}} 
\newcommand{\beqa}{\begin{eqnarray}} 
\newcommand{\eeqa}{\end{eqnarray}} 
\newcommand{\x}{S} 
\newcommand{\y}{{X_{+}}} 
\newcommand{\mb}{{X_{-}}} 
\newcommand{\ks}{{\delta_{GS}}} 
 
\begin{document} 

\pagestyle{empty} 
\begin{flushright} 
MPI-PhT-97-017 
\end{flushright} 
\vspace{1ex} 
 
\begin{center} {\large \bf  
 
Anomalous D-term, dynamical supersymmetry breaking and 
dynamical gauge couplings} 
      \rm 
\vspace{5ex} 
 
{\bf Z. Lalak$^{\S,\ddagger}$} 

\vspace{0.2cm} 
 $^{\S}$ Institute of Theoretical Physics \\ 
       {\em University of Warsaw} \\ 
       {\em 00-681 Warsaw}  

\vspace{0.2cm} 
 $^\ddagger$ Max Planck Institut f\"ur Physik \\ 
       {\em Heisenberg Institute}\\ 
       {\em D--80805 M\"unchen, Germany} 
 
\vspace{2ex} 
 
ABSTRACT 
\end{center} 
We analyze the structure of the vacuum and supersymmetry breaking  
pattern in Fayet-Iliopoulos models with dynamical gauge coupling  
and planck-scale value of the F-I parameter. 
We show that in this class of models supersymmetry  
is generically broken, but the mere presence of the D-term is not  
sufficient to 
stop the running away of the modulus responsible for the value of the  
gauge coupling - the dilaton.  
To stabilize the dilaton, one has to include an additional dilaton-dependent  
part in the superpotential. The presence of the large D-term gives rise   
to the mixed dilaton/D-term  
dominated scenarios of susy breaking, which allow  
horizontal hierarchy generation.
Models which can serve as secluded sectors in gauge mediation 
scenarios are discussed. It is shown  that when the F-I parameter and the 
gauge coupling are dynamical 
variables, the D-term dominated Universe does not allow for an 
inflationary period.  
 
\vspace{2ex} 
%
 
%
\newpage 

\pagestyle{plain} 
\section{Introduction} 
 
 
One of the main problems of unification within the framework 
of string-derived effective field theories is that of 
hierarchical supersymmetry breaking and the determination of 
the expectation value of the dilaton at a suitable scale. 
 
The dilaton puzzle has been reviewed in many papers, 
\cite{b1}, and is due to the fact that simplest 
nonperturbatively induced potentials for the dilaton don't 
seem to be able to stabilize it in a phenomenologically 
relevant region. While the more sophisticated versions of stringy 
unification, like $M$-theory, are under development which shall  
perhaps fix the dynamical gauge coupling in a fundamental way,  
we would like to explore in greater detail possibilities offered by the  
standard supergravity models.   
 
{}From the field theoretical point of view there are in fact  
two independent 
``experimental'' constraints on the dilaton expectation 
value: the requirement that the soft mass parameters which parametrize 
the observable breaking of supersymmetry be in the TeV 
region, and that the gauge coupling at the string scale be 
compatible with estimates based on renormalization 
group evolution of the standard model gauge couplings. 
The relationship between the gravitino mass and the  
dilaton vev involves the beta function coefficient of the 
strongly interacting hidden sector and thus depends on the 
possible matter content of that sector.  
 
In models derived from the heterotic string at tree level  
the universal gauge coupling constant $g_{string}$ 
is determined by the vev of the dilaton field \cite{b2} 
via  
\beq 
S=\frac{4 \pi}{g_{string}^2} - i \frac{\theta}{2 \pi} 
\label{s20}   
\eeq  
This normalization of the dilaton is such that under 
$\theta\rightarrow\theta+2\pi$ one has $iS\rightarrow iS+1$.  
There are nonuniversalities which can appear at the one-loop 
level and depend on remaining moduli fields $T,...,U$ 
\cite{b11}. It follows that the vevs of the dilaton and 
other moduli should determine gauge couplings in the hidden and 
in the observable sector. In particular, correct values of the QCD 
coupling $\alpha_s$ and the weak mixing angle $\sin^2 
\theta_W$ should follow. These requirements put strong 
constraints on the expectation value of the dilaton and on 
$\alpha_{string}$, i.e. one needs 
\beq \alpha_{string} = 
\frac{g^2_{string}}{4 \pi} \approx \frac{1}{20}  
\eeq  
In the most ``realistic'' scenarios, where a potential for the dilaton 
gets generated through gaugino condensation in the hidden sector one 
needs several condensates, or new symmetries and/or corrections to the 
K\"ahler potential of the dilaton in order to produce a minimum in the 
dilaton potential away from $0$ and $\infty$. Even then, in the case 
of models motivated by S-duality the dilaton gets naturally stabilized 
at a value of the order of unity, clearly outside the favoured  
region. A similar situation arises in multigaugino condensate models, 
where the dilaton vev is typically larger than $1$, but still  
smaller than $20$. Also, the question of supersymmetry breaking at 
that nontrivial minimum for the dilaton is a subtle one \cite{b1}. 
 
Once a dilaton potential appears, it is natural to ask what the mass 
of the dilaton is. It is important that the dilaton should receive a 
mass which is large enough to avoid cosmological problems, and to 
justify the treatment of the dilaton as a fixed background in low 
energy models, in other words to suppress the fluctuations around  
the dilaton background. This requirement is rather  
hard to fulfill. In the hidden sector susy breaking scenarios 
soft supersymmetry breaking masses vanish in the limit $M_{pl} \rightarrow  
\infty$, and the natural scale for the moduli fields is $M_{pl}$ itself, 
hence the usual sistuation is $m_s \ll <s>$, cosmologically dangerous.  
  
The typical scenario which one has in mind while discussing 
supersymmetry breaking and fixing the moduli vaccuum 
expectation values relies on the assumption that one can 
separate from the whole model the sector containing the 
dilaton and moduli, minimize it on its own, and then 
substitute fixed vevs of $S,T,U,...$ into the lagrangian 
describing remaining fields, which will have in turn their 
own separate dynamics. One always assumes, that the vevs of 
dilaton and moduli stay frozen at their values obtained at 
the first step of the above procedure, independently of what 
is happening in the chiral-gauge sector of the model. This 
is sometimes justified by the hierarchy of scales in a given 
model, however in general this point of view doesn't have to 
be correct. In particular, the backreaction of the other 
fields on the dilaton (moduli) can be significant in a class 
of models where there are from the beginning large terms in 
the lagrangian which contain both moduli and non-moduli fields.  
This is precisely 
the case in models with the anomalous $U(1)$ and the 
associated D-term, which is generically of the order of the 
Planck scale, $V_D \approx M^4_{planck}$.  
With the appearance of the large D-term there open up  
naturally new possibilities in the supersymmetry breaking mass  
patterns. Indeed, with the low energy effective lagrangian taking 
 the form 
\beq 
L_{eff}= Z_{i \bar{j}} \partial \phi^i \partial \phi^{\bar{j}}  
- m^2_{i \bar{j}} \phi^i \phi^{\bar{j}} + ... - g^2 (\xi + 
Z_{i \bar{j}} \phi^i \phi^{\bar{j}} q_i )^2 
\eeq 
the masses that are softly breaking supersymmetry\footnote{ 
To identify the physically meaningful soft terms we assume 
 that at the minimum the cosmological constant vanishes in the  
underlying supergravity model} have the F-term  
 contributions \cite{b4} 
\beq 
\delta_f m^2_{i \bar{j}} = m^2_{3/2}  
Z_{i \bar{j}} - F^{\alpha} F^{\bar{\beta}}  
 R_{\alpha \bar{\beta} i \bar{j}} +  
m^2_{3/2} \nabla_i G_k \nabla_{\bar{j}} G^k
\eeq 
where $R_{\alpha \bar{\beta} i \bar{j}}$ is the curvature tensor of the 
K\"ahler scalar manifold,  
and the D-term contributions  
\beqa 
\delta_d m^2_{i \bar{j}}& =& g^2 G_{\bar{j}} G_i D^2 -2 g^2 D 
(G_i D_{\bar{j}}  
+ D_i G_{\bar{j}}) + \nonumber \\ 
 & + & 2 g^2 D_i D_{\bar{j}} + 2 g^2 D D_{i \bar{j}} + \nonumber \\ 
 & - &  g^2 Z_{i \bar{j}} D^2  
\label{mds} 
\eeqa 
The (\ref{mds}) is in general nonuniversal, however it simplifies  
if indices $i,\bar{j}$ correspond to non-messenger fields, i.e. to the  
fields which do not lie along susy-breaking directions. In this simpler 
case the (\ref{mds}) reduces to ($M=M_{pl}$)  
\beq 
\delta_d m^2_{i \bar{j}} = 
2 g^2 Z_{i \bar{j}} q_i <D> - g^2 Z_{i \bar{j}} \frac{<D^2>}{M^2} 
\label{simp} 
\eeq 
Of course, on the rhs of this formulae the first contribution  
dominates over the second, however the second contribution can in specific  
cases be comparable to  $\delta_f m^2_{i \bar{j}}$.  
The omitted terms are important for messengers, the fields which are usually   
the heaviest ones among the non-moduli fields, however it is the nonmessenger  
fields, among them hopefully the MSSM fields, which are of immediate interest  
to us.   
For non-messengers the F-type contribution tends to be  
nonuniversal, but the simplified D-type 
contribution (\ref{simp}) exhibits the natural alignment, and if the D-term  
is sufficiently large, the aligned terms can dominate soft masses for  
$U(1)$ charged fields supplying the required amount of universality.  
Also, in that case, the charged states tend to be heavier than  
the un-charged ones which can be seen as a source of hierarchy.  
It should be noted, that the D-type contributions depend, explicitely 
through the $g^2$ and implicitely through the vev of $D$, on the  
dilaton.  
 
It is obvious then, that the large F-I term can not only affect  
the physics of the supersymmetry breaking sector, but also has  
direct impact on the low-energy effective theory. 
 
In this paper we want to discuss general features of  
dilaton stabilization and susy breaking in the presence of the  
F-I term and non-perturbative superpotential for messengers.

\section{The dilaton potential in the presence of perturbative superpotential 
for $U(1)$ charged fields} 
\label{sec:ic} 
\vspace{0.5cm} 
 
To start with we will focus on the part of the effective 
action which involves only the dilaton and a set of fields 
$X_i$, charged under an anomalous $U(1)$.  
Let's consider the low-energy supersymmetric implementation of  
the stringy anomalous $U(1)$ under which the dilaton superfield S  
gets shifted by a  
chiral superfield parameter. Such a shift in the universal gauge kinetic  
function induces a term which corresponds to a mixed anomaly which includes  
the  
$U(1)$ gauge field. To cancel this, we need matter charged both under $U(1)$  
and under all other factors of the nonanomalous gauge group, including the  
strongly coupled hidden sector group. This implies that in principle we  
shall have hidden matter, and matter superfields condensates, denoted later  
by T, which should be  
taken into account when writing down the effective lagrangian. 
 
The K\"ahler potential is \cite{b5} 
\beq 
K= - M^2 \log (S/M+\bar{S}/M + \delta_{GS} V/M^2) + \sum_i  |X_i|^2 
\eeq 
 
Here $V$ is the vector superfield of the anomalous $U(1)$, canonical  
kinetic terms are assumed.  
The relevant F-terms which control the effects of supersymmetry 
breaking are  
\beq 
F^S =  e^{\frac{K}{2 M^2}} (W_S - \frac{W}{S+\bar{S}}) \frac{(S + \bar{S})^2}
{M^2}
\eeq 
and  
\beq 
F^{X_i} = e^{\frac{K}{2 M^2}} (W_{X_i} + \frac{W \bar{X}_i}{M^2}) 
\eeq 
Finally, the scalar potential including the anomalous D-term contribution 
is  
\beqa V &=& \frac{1}{M^2} e^{\frac{K}{M^2}} (|(S+\bar{S} )W_S -W|^2 + 
\sum_i |M W_{X_i} + \frac{W \bar{X_i}}{M}|^2 \nonumber \\ &-& 3 |W|^2 ) + 
\frac{4 \pi M^5}{S+\bar{S}} (\frac{8 \pi \delta_{GS} M}{S+\bar{S}} +  
\frac{1}{M^2} \sum_i 
q_i |X_i|^2)^2  
\eeqa 
where in the case of the stringy model we have $\delta_{GS} = \frac{  
TrQ }{192 \pi^2}$. 
As usual, we assume that there exist fields with negative 
charges with respect to the anomalous $U(1)$ so that the 
D-term can vanish for some field configuration.  
It is also obvious, that we need a nontrivial superpotential for  
X-fields. Let us assume for a moment that there exist a field, 
$B$, which has no superpotential interactions at all, and is charged 
only under the anomalous $U(1)$. Then the relevant part of the potential
is 
\beq
V_B = \frac{4 \pi M^5}{S+\bar{S}} (\frac{8 \pi \delta_{GS} M}{S+\bar{S}} -   
\frac{1}{M^2} 
|q_B| |B|^2)^2
\eeq
The eigenvalues of the mass matrix given by this potential are: 
$0$ for the combination $\phi_0 = -\sqrt{\frac{|q_B|}{\pi \delta_{GS}}} 
(<s_r>/M)^{3/2}\; \delta s_r\;  + \; \delta b$ of the fluctuations of the 
fields around the spontaneously chosen vacuum, and $\sim M^2$ for the 
orthogonal combination $\phi_m$. After proper normalization the field $\phi_m$ 
becomes the scalar superpartner of the massive vector multiplet $V$ which 
contains the gauge boson of the anomalous $U(1)$ with the combination 
of $s_I$ and $\theta$ - the phase of the field $B$ - as its longitudinal component. 
The mass of the complete massive vector supermultiplet which decouples in a 
supersymmetric way from the low-energy model is $g^2 M$.
The orthogonal combination $\phi_0$ is massless, which means that the $g^2$ 
is left  undetermined. In this case, which happen to occur in known stringy 
models and M-theory models, the Fayet-Iliopoulos term plays no role for the low-energy theory.  \\
In reality the potential given by the anomalous D-term is entered  
by a larger number of  fields (dilaton and at least charged chiral fields necessary  
to cancel anomalies in both hidden and visible sectors) 
 and has flat directions, 
which precludes dilaton stabilization.

However, one can assume the point of view that  the presence of the 
singlet $B$ is not really guaranteed for all possible string or M-theory 
compactifications, and continue the investigation under the assumption that 
it is meaningful to ignore such singlets. This is rather strong assumption, but
it can be justified by the  inetersting phenomenology of the models of the 
Fayet-Iliopoulos 
type \cite{pokor}. It implies that we agree to tolerate vacua with large 
D-terms as long as the effective scale of these D-terms is hierarchically 
smaller than the planck scale.
     
Further to the above assumption one can in imagine, that one 
constructs a perturbative  
superpotential for Xs which together with some generic nonperturbative  
superpotential for S, and perhaps together with gravitational corrections,  
would fix the dilaton, and this way produce supersymmetry  
breaking\footnote{Of course one always needs a 
nonperturbative part of the superpotential which  
contains S - one knows that purely perturbative effects cannot break  
supersymmetry in stringy models}.  
 
There are two obvious problems with the idea of using purely perturbative  
superpotential in the X-sector. 
First, there are in fact many chiral superfields in the model, and as usual  
in O'Reiferteigh type models one expects many flat directions which  
generically upset the dilaton stabilization in the manner based on the D-term, 
and second, and most important, it turns out that we need a new mass scale, 
say  $M_I$, about 2 orders of magnitude below the planck scale, 
in the X-sector. The most economic and interesting possibility is that  
the new scale is in fact related to the hidden sector condensation.  
This scenario is discussed in the remaining part of this paper.

\section{Fayet-Iliopoulos model with the dynamical mass scale} 
 
Let's consider more detaily the low-energy supersymmetric implementation of  
the stringy anomalous $U(1)$. 
 
To procede assume  that there are fields $G_{-}, \; G_{+}$ charged 
under $U(1)$ 
and under condensing $SU(N)$ in $\bar{N}$ and $N$ representations  
respectively (just one pair  
of  $\bar{N}$ and $N$). 
 Let's call another  pair  
of  
fields, charged under anomalous $U(1)$ but singlet under $SU(N)$, $X_{-}$ and  
 $X_{+}$ - these are the would-be messengers, by which we mean 
 the fields whose F-terms  
can take nonzero expectation values, and which can couple to the observable  
sector through superpotential couplings.  
There exists an allowed by symmetries perturbative superpotential  
coupling of these fields  
\beq 
W_{pert} = \frac{G_{-} G_{+} X_{-} X_{+}}{M} \lambda 
\label{pert} 
\eeq 
(We take $\bar{q} = Q(G_{-})=Q(X_{-})=q_{-}=-1$ and define $ q = Q(G_{+})$,  
$q = Q(X_{+})$, the obvious generalization shall be discussed later).  
In the presence of matter the nonperturbative superpotential dictated  
by nonanomalous symmetries is, cf \cite{b8} and forthcoming sections of this letter, 
\beq 
W_{npert}= U \log \left ( \frac{U^{N-1} \det T}{\Lambda^{3 N -1}} \right ) - U (N-1) 
\eeq 
where $U$ is gaugino condensate superfield and $T=G_{-} G_{+}$.  
The $W_{pert}$ can be treated as an supersymmetric mass term for $ G_{-}  
G_{+}$. 
 
Assuming, as in previous section, that supersymmetry is not broken   
along the condensate directions,  
one can integrate out from $W_{pert}+W_{npert}$ the superfields $U$ and  
$T$ obtaining an effective  
superpotential for messengers $ X_{-}$ and $ X_{+}$ 
 
\beq 
W_{eff} = \Lambda^3 \left ( \frac{ X_{-} X_{+} }{ M \Lambda} \right )^{\frac{1}{N}}  
\label{weff} 
\eeq 
 
with  
\beqa  
<U> & = &  \Lambda^3 \left ( \lambda \frac{ X_{-} X_{+} }{ M \Lambda} \right )^{\frac{1}{N}} \nonumber \\ 
<T> & = &   \Lambda^2 \left ( \lambda 
\frac{ X_{-} X_{+} }{ M \Lambda} \right )^{\frac{1-N}{N}}  
\eeqa 
 
One can build the potential for $  X_{-}$ and $ X_{+} $ using the  
effective superpotential (\ref{weff}) and known form of the  
D-term contribution 
\beq 
V_D = \frac{g^2}{2} (q_{+} |X_{+}|^2 -|X_{-}|^2 + \xi)^2  
\eeq 
Keeping for a while $g^2$ and $\xi$ as constant parameters 
one arrives at equations of motion for Xs. It is straightforward to show  
that these equations have no solution for any value of $N$, $\xi$ and 
 $g^2$, 
in sharp contrast to the usual Fayet-Illiopoulos model \cite{b6} with perturbative  
supersymmetric mass analyzed in the paper \cite{b7}.  
The point is that in fact supersymmetry is broken along the $T$ direction 
in this type of models, so one is not allowed to integrate out $T$ using 
its supersymmetric equations of motion. This is an interesting point,  
as at the end it turns out that $F_T$ is subdominant with respect to other 
$F$-terms thorought the parameter space, however, approximating it  
by zero prevents one from finding the  
true ground state.  
  
Next step is to make $S$, the dilaton, a dynamical degree of freedom. 
Using our conventions, $Re(S) = \frac{4 \pi}{g^2}$, 
and noticing that $\Lambda = e^{- \frac{ 2 \pi}{b_o} S_{r}}$ where  
$b_o = 3 N -1$  
one obtains the effective superpotential for $S,   X_{-}$ and $ X_{+} $ 
in the form 
\beq 
W= M^3 e^{-\frac{2 \pi S}{ N }}  
\left ( \frac{ X_{-} X_{+} }{ M^2} \right )^{\frac{1}{N}}  
\eeq 
 
and the D-term contribution with variable $S$ and explicit powers of  
Planck scale M is  
\beq 
V_D = \frac{M^4 \; 4 \pi}{S+\bar{S}} (\frac{8 \pi \delta_{GS} M }{S+\bar{S}} +  
\frac{1}{M^2} (q_{+} |X_{+}|^2 -|X_{-}|^2))^2 
\eeq 
  
This superpotential together with the D-term contribution leads to 
a rather complicated set of equations. However, the numerical study  
of these equations hints that there appears the standard runaway  
behaviour in the direction of $S_r$, well known from  
single-gaugino-condensate models. The next step is  
to go to the full, $T$-dependent superpotential.  
 
Now, the term  
\beq 
\delta W = \lambda \frac{ T X_{-} X_{+}}{M} 
\label{mix} 
\eeq 
is not required by first principles, so one should check first  
what happens with the simple superpotential  
\beq 
W= M^5 \frac{e^{-\frac{10 \pi}{3 N -1} S}}{T} 
\eeq 
The answer is that there is the run-away behaviour with $y\rightarrow 0,  
X_{-} \approx 1/\sqrt{Re(S)}, X\rightarrow \infty, s \rightarrow \infty$,  
both in the global and in the local case. Hence,  
the  term (\ref{mix}), which plays the role of the supersymmetric  
linear term for $T$, known from non-F-I dynamical supersymmetry  
braking models - cf. \cite{b8}, has to be included.  
Then the situation is at first sight not that obvious. In globally  
supersymmetric case one can easily convince oneself that there is  
runaway behaviour also in the presence of (\ref{mix}). But in the local case, 
the form of the potential hints at the possibility that  
$T$ stabilizes at $M^2$, which - while $  X_{-} X_{+}$ is nonzero 
at the minimum - would imply that $S$ gets stabilized  
like in gaugino condensate models with a c-number constant in the superpotential (the constant being played by a nonvanishing vev of (\ref{mix})).  
However, this scenario doesn't seem to be realized in a simplest model  
with the simplest hidden-visible mixing term (\ref{mix}). What happens in this  
particular model is again the run-away behaviour, largely because of the presence of the additional degree of freedom $T$. In fact, the role of the field $T$  
is similar to that played by the ``radial'' modulus T in heterotic string  
compactifications, with the crucial difference that in the stringy case  
the T-modulus has a strong stabilizing superpotential proportional  
to $1/ \eta^6 (T)$. 
 
The conclusion of this section is that simple dynamical  
scenarios which fail to stabilize the dilaton in non-F-I models, 
fail to do so also in the presence of the large, stringy, F-I term.

\section{More general F-I models with dynamical supersymmetry breaking
- secluded sector models}

So far we have discussed  a simple SU(2) model with $N_f=1$. Let 
us consider in some detail more general models with $N_f < N$. 
The nonperturbative superpotential is 
\beq
W= (N-N_f) (\frac{\Lambda^{b_o}}{\det T})^{1/(N-N_f)}
\eeq
where $b_o=3 N - N_f$ and the matrix $T$ is defined as $T_{ij}= Q_i \tilde{Q_j}$.
where  $\tilde{Q_j}$ transforms as $\bar{N}$. We assume the canonical 
kinetic terms for the matter superfields. In addition to the F-type potential 
coming from $W$ one has also nonabelian D-terms of the group $SU(N)$ and the 
Fayet-Iliopoulos D-term of $U(1)$. In first step we look for the solutions which make the nonabelian D-terms vanish. This is achieved by taking the following form of 
$T$: $T_{ij} = \delta_{ij} |v_i|^2$ with $i=1,..., N_f$. the scalar potential along these directions is 
\beq
V= \frac{ 2 \Lambda^{2 b_o /(N-N_f) }}{\prod_i |v_i|^{4/(N-N_f)}} 
( \sum_j \frac{1}{|v_j|^2} ) + \frac{g^2}{2} (\xi + \sum_k (q_k + \bar{q}_k)
|v_k|^2)^2
\label{gpot}
\eeq
Let us look for a symmetric solution of this potential: $|v_i|^2 = |v|^2 = x$,
 $q_k + \bar{q}_k= - 2 \pi \delta_{GS}$, $\xi = g^2 M^2 \delta_{GS} N_f$. 
With these assumptions one gets the simpler potential
\beq
V= \frac{ 2 N_f  \Lambda^{2 b_o /(N-N_f) }}{x^{2 N_f /(N-N_f)}} 
 + \frac{g^2}{2} (\xi - 2 \pi \delta_{GS} N_f x)^2
\eeq
This potential can be easily minimized for $x$ giving 
\beq
x= \frac{g^2 M}{2 \pi} (1 + \delta), \; \delta \approx (\Lambda^2 / M^2 )^{
\frac{3 N - N_f}{N-N_f}}
\label{soln}
\eeq
This solution corresponds to 
\beq
F_Q \approx \Lambda^2 (\frac{\Lambda}{M})^{\frac{N+N_f}{N-N_f} },\;
D \approx M^2 (\frac{\Lambda^2}{M^2})^{\frac{3 N - N_f}{N-N_f} }
\label{fd}
\eeq
From (\ref{fd}) one can see that $F^2_Q \approx M^2 D$ which implies 
that $F_Q$ is much larger than $D$, however, the contribution of both sources 
to the soft masses is  similar
\beq
\delta_f m^2 \sim F^2/M^2 \sim D \sim \delta_d m^2 
\eeq
One also finds the general upper limit on the soft masses generated so this way 
valid for any allowed $N, N_f$:
 $\delta m^2 < \Lambda^2 ( \Lambda / M )^2 $. It is easy to check  that the actual 
magnitude of these masses falls down quickly with growing  $N, N_f$.
At this point we can assume in the spirit of the gauge mediated 
supersymmetry breaking models that there exist an absolute gauge  singlet field 
$Z$ which couples at the renormalizable level to the hidden sector, 
and also to the messenger sector. This possibility is of particular
interest in the hypothetical models where the observable matter does not 
carry the F-I $U(1)$ quantum numbers. 
For simplicity let us assume that the singlet $Z$ couples to the first 
generation of matter fields
\beq
\delta W = \lambda Q_1 \tilde{Q}_1 Z
\eeq
Then we have to single out the vev of the first generation and call its 
square $x_1$ (and the common value of the squared vevs of remaining generations
$x$ as previously). The important new terms in the scalar potential 
are
\beq
V = ... + x_1 | \lambda z - \Lambda^{\tilde{b}_o } x^{-\frac{N_f-1}{N-1} -
\frac{N - N_f +1}{N-N_f}}|^2 + |\lambda|^2 |x_1|^2 + ...
\label{gcor}
\eeq
The solution to the potential can be obtained also in this case, 
and it gives 
\beq
F_{Q_1} \approx 0, \; F_{Q_{i \neq 1}} \neq 0 \;(and \;\; dominant)
\eeq
and 
\beq
|F_Z| \approx |v_1|^2 \neq 0, \; |z| \approx  |v_1|, \; |v_1|^2 \ll  |v|^2
 \sim g^2 M^2 
\eeq
These properties of $F_Z$ and $Z$ show that the hidden sector model discussed here
could serve as a rather simple ``secluded'' sector in scenarios with gauge 
mediation of the supersymmetry breaking between secluded  and observable sectors. It is interesting to note at this point that in the potential 
(\ref{gpot},\ref{gcor}) the D-term plays the role analogous to the soft mass terms in explicitely broken SQCD. It induces finite condensates 
$ < \psi_{Q_i} \tilde{\psi}_{\tilde{Q}_j} > \sim < F_{T_{ij}} > \neq 0$
although the breaking of supersymmetry is only spontaneous here.

However, one should ask the question what happens in these models 
when the gauge 
coupling squared becomes the inverse of the dynamical field $S$. 
Then the potential computed above obtains a new contribution 
\beq
V \rightarrow V + |\frac{\partial W}{\partial S}|^2  \sim 
 \frac{v^2}{M^2}  |F_Q|^2 + \frac{|F_Q|^4}{M^4}
\eeq
where the second term is the new F-term, taken at the minimum, and the last 
term comes from the D-term. As expected the $F_Q$  falls down exponentially 
with $S$ and there is the run-away behaviour towards the ultra-weak 
coupling region. The situation is not improved when one takes into account 
gravitational corrections in the full supergravity lagrangian.   
Hence, the negative conclusion about the stabilization of the 
dynamical gauge coupling in the presence of the large F-I term remains valid. 

Then the question is what happens to the models which are known  
to give reasonable results without the D-term, like race-track models, 
cf. \cite{b9}, or S-dual models of \cite{b1}. 

\section{Dynamical F-I models with superpotential which stabilizes the dilaton} 
 
The general answer to the question posed at the end of the previous section 
is that these models work, i.e. stabilize the dilaton, also in the F-I case,  
but their predictions get affected.  
 
To illustrate the role of the D-term in a model independent way  
we shall discuss two toy models which  
however are designed to resemble the typical situations encountered  
in popular models with dynamical gauge coupling: a) Model I, which  
corresponds to the situation where the superpotential itself stabilizes  
the dilaton at some scale equal or larger than the planck (or string) scale 
(like in S-dual models of \cite{b1}), 
b) Model II, where superpotential alone stabilizes the dilaton at some  
unacceptably small scale (which can easily happen in race-track models). 
We shall consider separately the two cases.  
  
\noindent{\bf model I} 
 
We consider the following model 
\beq 
K=- M^2 \log (\frac{\x +\bar{\x}}{M}) + \y \bar{\y} + {\mb}\bar{\mb} 
\eeq 
\beq 
W= m {\mb}\y + q (\x-p)^2 
\eeq 
where  $\x$ is the dilaton superfield, and $\y$ and $\mb$ 
are ``messenger'' superfields, charged under $U(1)$.  
We assume everywhere that $m \leq q \; \ll M \;(=M_{pl})$ 
but, as required by gauge coupling unification, $p>M$\footnote{This  
superpotential should be regarded as a part of a  
series expansion, cf. Section 7.}.  
 
If we put aside the charged fields and the D-term, the model has its  
minimum at $\x =p$, and there is another minimum of the potential at  
$\x + \bar{\x}=0$, but with present choice of the K\"ahler function it is  
infinitely far away in the field space. In the presence of the D-term, the  
situation changes. The minimum at small $x$ appears (comes in from infinity) 
at $x_{sm}=\frac{M}{2} (\ks/(8 \pi^2 \; m^2/q^2 \; M^2/p^2)^{1/3}$. Under   
our assumptions this tends to be  smaller than $M$, hence doesn't  
correspond to  
the phenomenologically required solution, 
unless $m$ becomes comparable with $q$ . 
At this point  
supersymmetry is broken with $F_{\x} \gg <D>$, cf. model II in  
the next subsection.  
The second minimum appears in the vicinity of 
$p$. 
 
The assumed form of the K\"ahler potential and of the 
superpotential gives the scalar potential  
\beq 
V= m^2 (|\y|^2 +|\mb|^2) + \frac{(\x + \bar{\x})^2}{M^2} 4 q^2 
|\x-p|^2 + \frac{4 \pi M^5}{(\x +\bar{\x})} \left (\frac{8 \pi \ks M}{(\x 
+\bar{\x})} - \frac{|\mb|^2}{M^2} +q_{+} \frac{|\y|^2}{M^2} \right )^2  
\eeq 
(we assume $\ks \approx 0.01$, as given by a typical string model, 
and $q_{+} = 2 \pi \ks +1$ - as explained in Section 7.) 
 
It is more or less obvious that this potential can have a 
minimum at finite $\x,\y,\mb$. Also, it is clear that if there is a 
minimum, it has to correspond to $\y=0$. Assuming further that  
we can restrict ourselves to real values of the scalar components of 
superfields, we shall solve perturbatively  
equations  
\beq 
\frac{\partial V}{\partial \y}=0, \; \frac{\partial V}{\partial 
\mb}=0, \; \frac{\partial V}{\partial \x}=0 
\eeq 
 
From $\frac{\partial V}{\partial 
\mb}=0$ we get  
\beq 
\mb^2 = \frac{1}{2 M} ( \frac{8 \pi \ks M^4}{\x} -m^2 \frac{\x}{2 \pi})  
\label{5} 
\eeq 
or $\mb=0$. The latter possibility cannot correspond to minimum  
(D-term would be large) so we take the first possibility.  
 
From $\frac{\partial V}{\partial \x}=0$ we obtain, after substituting  
(\ref{5}), the equation for $\x$ which we can solve perturbatively. 
To this end we assume $\x=p (1+ e)$. Then in the leading order we get  
\beq 
e= \frac{\pi}{8} \ks \frac{m^2}{q^2} \frac{M^5}{p^5} \; (<<1) 
\label{6} 
\eeq 
 
With this we can compute the F-terms: 
\beqa 
|F_{\mb}|&=& 0 \\ 
|F_\y|&=& m M \sqrt{\ks \frac{M}{p} 4 \pi} \\ 
|F_\x|&=& |(K_{\x \bar{\x}})^{-1} W_\x |= m M \frac{\ks}{2} \frac{M^2}{p^2}  
\frac{m}{q} 
\eeqa 
 
and we can find the contribution to soft scalar masses given by the  
nonvanishing D-term 
\beq 
\delta m_i^2 = q_i g^2 2 <D> = q_i m^2  
\eeq 
(this is a contribution for the field $X_i$ with the charge $q_i$.) 
 
In this case the $F_{\x}$ is nonzero, but smaller than $F_{\y}$, 
the F-term corresponding to positively charged messenger, unless  
$m$ is comparable with $q$. In general, it is the D-term contribution  
which is dominant in the soft susy breaking mass parameter of the  
charged scalars, $(F_{\y}/M)^2 = 4 \pi m^2 (\ks M /p) < m^2$.

\vspace{1cm} 
 
\noindent{\bf model II} 
 
This is the model which in the absence of the $U(1)$ and its  
D-term correspond to strongly coupled vacuum.  
However, the presence of the planck-scale D-term changes  
situation dramatically. As before the K\"ahler function is  
\beq 
K=- M^2 \log (\frac{\x +\bar{\x}}{M}) + \y \bar{\y} + {\mb}\bar{\mb} 
\eeq 
and the superpotential 
\beq 
W= m {\mb}\y + q  \x^2 
\eeq 
In these formulae $\x$ is the dilaton superfield, and $\y$ and $\mb$ 
are ``messenger'' superfields, charged under $U(1)$ - as previously.  
We assume again  that $m, \; q \; \ll M $.  
The scalar potential is  
\beq 
V= m^2 (|\y|^2 +|\mb|^2) + \frac{(\x + \bar{\x})^2}{M^2} 4 q^2 
|\x|^2 + \frac{4 \pi M^5}{(\x +\bar{\x})} \left (\frac{8 \pi \ks M}{(\x 
+\bar{\x})} - \frac{|\mb|^2}{M^2} +q_{+} \frac{|\y|^2}{M^2} \right )^2  
\eeq 
(we assume $\ks\approx 0.01$, as given by a typical string model, 
and $q_{+} = 2 \pi \ks +1$ - as explained in Section 7.) 
 
It is more or less obvious that this potential can have a 
minimum at finite $\x,\y,\mb$. Also, it is clear that if there is a 
minimum, it has to correspond to $\y=0$. Assuming further that  
we can restrict ourselves to real values of the scalar components of 
superfields, we shall solve perturbatively  
equations  
\beq 
\frac{\partial V}{\partial \y}=0, \; \frac{\partial V}{\partial 
\mb}=0, \; \frac{\partial V}{\partial \x}=0 
\eeq 
 
From $\frac{\partial V}{\partial 
\mb}=0$ we get again 
\beq 
\mb^2 = \frac{1}{2 M} ( \frac{8 \pi \ks M^4}{\x}  
-m^2 \frac{\x}{2 \pi})  
\label{2.5} 
\eeq 
From $\frac{\partial V}{\partial \x}=0$ we obtain, after substituting  
(\ref{2.5}), the equation for $\x$  
\beq 
-\frac{m^4}{4 M} - 8 \pi^2  
\frac{\ks M^3 m^2}{\x^2} + 2 \pi \frac{64 q^2 \x^3}{M^2} =0 
\label{2.6} 
\eeq 
 
This we can solve perturbatively assuming $m^2, q^2 <<M^2$. 
Knowing that we should get $\x \approx M$ we shall retain in the leading order  
only last two terms on the LHS of (\ref{2.6}) and solve for $\x_o$ 
\beq 
\x_o = (\frac{\ks m^2}{2^9 \pi^4  q^2})^{1/5} M 
\eeq 
To procede we  assume $\x=\x_o (1+ e)$. Then in the leading order we get  
\beq 
 \x = (\frac{\ks m^2}{2^9 \pi^4  
 q^2})^{1/5} M (1 + \frac{m^2 (\ks m^2 /q^2/(8 \pi^4))^{2/5}}{ 
2^{2/5} (160 \ks M^2 -2^{3/5} m^2 (\ks m^2/q^2/(8 \pi^4) )^{2/5})}) 
\label{2.7} 
\eeq 
 
If we substitute this result into (\ref{2.5}), we obtain the leading order  
value of $\mb$.  
 
With this we can compute the F-terms: 
\beqa 
|F_{\mb}|&=& 0 \\ 
|F_\y|&=& m ( (\frac{\ks m^2}{ 8 \pi^4 q^2})^{1/5} (2560 \ks M^4 (\frac{\ks m^2} 
{8 \pi^4 q^2})^{3/5} 
q^2 16 \pi^4 \nonumber \\ 
 & -& 208 \;  2^{3/5} \ks m^4 M^2))^{1/2} / (2^{4/5} 640 \ks m^2 M^2)^{1/2}\\ 
 & =& 2 m M (\ks)^{2/5} (\frac{q}{m})^{1/5} (4 \pi^2)^(1/5) \\ 
|F_\x|&=& M q (\ks)^{3/5} (\frac{m}{q})^{6/5} \frac{1}{(4 \pi^2)^(1/5)} 
\eeqa 
It is to be noted that because of the nontrivial K\"ahler function  
there is the relation $F_\x = \frac{W_\x}{K_{S \bar{S}}} = 4 \frac{\x^2}{M^2} 
W_\x$.  
In the next step we can find the  
contribution to soft scalar masses given by the  
nonvanishing D-term (notation as before) 
\beq 
\delta m_i^2 = q_i g^2 2 <D> \approx  q_i m^2  
\eeq 
(this is the contribution for the field $X_i$ with the charge $q_i$.) 
In this case one can easily obtain required value of $s$ taking $m$  
suitably larger than $q$ (we didn't assume anything about the ratio  
$m/q$ to get the solution).  
As for the soft terms,  
\beq 
\frac{|F_{\x}|}{|F_{\y}|} = \frac{<s>}{ 2 \pi M} 
\eeq 
and  
\beq 
\frac{\delta_f m^2}{\delta_d m^2} = \ks \frac{<s>}{\pi M} 
\eeq 
Hence, if the vacuum in this model lies in the weak coupling regime, then 
the dilaton F-term can be as large as the other auxiliary fields, and  
its contribution to the soft masses can be comparable with 
the D-term contribution.  
This has various consequences. One is that the gaugino masses can be in this  
case as large as the scalar soft masses, which is very good from the  
point of view of low-energy phenomenology, the other consequence is,  
however, that the possibility of creating hierarchy through the assignement  
of $U(1)$ charges is essentially lost. 
 
One should note at this point, that we have been a bit cavalier about stringy  
properties of the effective models which we have discussed. In fact,  
in the stringy lagrangian all the terms, if we put F-I parameters to zero and  
omit truly nonperturbative terms, should be multiplied by the common power  
of $1/(S + \bar{S})$ as all of them come from the tree-level string amplitudes. 
This feature is not uniquely implementable into the globally supersymmetric  
lagrangian. We believe, that in principle the correct procedure of taking this  
feature into  
account is to minimize not the ``naive''  globally supersymmetric lagrangian, 
as we have been doing so far, but to minimize the leading part of the stringy  
locally supersymmetric lagrangian. The change in the potential would be that  
all the terms except the D-term would be multiplied  
by the factor $e^{K/M^2}$. Now, this factor is 
\beq 
e^{K/M^2} = \frac{M}{S + \bar{S}} ( 1 + \frac{1}{M^2} \sum_i |X_i|^2 
+ o(\frac{|X|^4}{M^4})) \approx \frac{M}{S + \bar{S}} 
\eeq 
in the leading order. The change in the equations which arises from this  
modification turns out to be inessential for the overall conclusions,  
although their actual form becomes more complex. Hence, to keep our discussion  
simple and quasi-analytical we  use the ``naive'' global  
models as the illustration.       
 
\section{Symmetries of the stringy hidden gauge sector in the presence of the  
$U(1)$} 
 
To narrow down the genarality of the discussion to the range of realistic models, 
let us discuss more carefully the possible dilaton-dependence of the  
effective superpotential in anomalous $U(1)$ models, in the manner based in  
symmetries.  
  
It is well known that the  supersymmetric $SU(N)$  theories with  
$N_f$ copies of $\bar{N} + N$ matter representations, $N_f \leq  N-1$, 
without dynamical coupling have  anomalous axial and R-symmetries. 
In the case discussed here, $N=2, N_f =1$ there are two independent  
anomalous symmetries: 
axial $U(1)_A$ ( with the associated Konishi current) under which  
\beq 
Q \rightarrow e^{i \alpha} Q, \; \bar{Q} \rightarrow e^{i \alpha} \bar{Q} 
\label{an1} 
\eeq 
and the R-symmetry $U(1)_R$      
\beq 
Q \rightarrow e^{2 i \alpha}  Q, \; \bar{Q} \rightarrow  e^{2 i \alpha} \bar{Q},\; 
V \rightarrow e^{- 6 i \alpha} V, \; \theta \rightarrow e^{3 i \alpha} \theta 
\label{an2} 
\eeq 
( this is the symmetry whose current lies in the supermultiplet 
with the stress tensor and the supercurrent). 
The linear combination of these two symmetries 
forms the nonanomalous R-symmetry 
$U(1)_{R'}$ 
\beq 
Q \rightarrow e^{-i \alpha} Q, \; \bar{Q} \rightarrow e^{-i \alpha} \bar{Q}, \; 
V \rightarrow V 
\label{r1p} 
\eeq 
One assumes that the nonanomalous R-symmetries should be respected by  
nonperturbative effects, \cite{b8}, and this way they constrain the  
form of the nonperturbative superpotential induced for low-energy gauge  
invariant degrees of freedom $T = Q \bar{Q}$ giving the well known form  
of the superpotential $\frac{1}{T}$. In the presence of dynamical  
gauge coupling, 
$L = ... \frac{1}{32 \pi } (S W^2 )_F + h.c.+...$ there  
appears a new anomalous  
symmetry which consists in an imaginary global shift of S 
\beq 
S \rightarrow S - i \alpha 
\label{an3} 
\eeq 
This again  
can be combined with the anomalous $U(1)_R$ to form a nonanomalous  
R-symmetry involving $S$ and $W^2$ superfields, cf. \cite{b10}. 
 
The question is whether 
the presence of the anomalous $U(1)$ in stringy models  
can restrict the possibilities. The point is that at the scale of the extra  
gauge boson mass that boson decouples, and below that scale we do not have  
at our disposal the gauge transformation which shifts S superfield.  
What is left after anomalous $U(1)$ is the global symmetry with charges equal  
to the charges under local $U(1)$. This is the classical invariance of the 
low energy matter lagrangian, but   
 it is anomalous. 
Since it is anomalous, we can combine it with the above global  
imaginary shift of $S$ to form a new nonanomalous global symmetry, which we  
could use to constrain the form of the Lagrangian. 
One can easily see that the nonanomalous combination is  
\beq 
S\rightarrow S-i \ks \alpha,\; Q \rightarrow e^{-i q \alpha} Q,\;  
\bar{Q} \rightarrow e^{-i \bar{q} \alpha} \bar{Q} 
\label{nac} 
\eeq 
where $q+\bar{q} = -2 \pi \ks$. 
It is easy to see that (\ref{nac}) gives the following general form of  
the superpotential 
\beq 
W = \left (\frac{e^{-2 \pi S}}{T} \right )^{\gamma} 
\label{wnac} 
\eeq 
It is obvious that to fix $\gamma$ we need the nonanomalous R-symmetry.  
Indeed, when one combines the imaginary shift with the  
R-symmetry (\ref{an2}), 
one obtains one more independent non-anomalous  
mixture of anomalous symmetries, which is actually an R-symmetry 
$U(1)_{R''}$ 
\beqa 
 S\rightarrow S- i \frac{\alpha}{\pi} (3 N - N_f),& 
 Q,\bar{Q} \rightarrow e^{2 i  \alpha} Q,\bar{Q}&  
V \rightarrow e^{- 6 i \alpha} V,
 \nonumber \\
& \theta \rightarrow e^{3 i \alpha} \theta&   
\label{r2p} 
\eeqa 
(we use $N=2, \; N_f=1$ in this section).
If we impose the symmetry (\ref{r2p}) on the effective superpotential  
containing just $S$ and $T$ then we obtain the general expression 
\beq 
W= M^3 e^{-2 \pi S} g(\frac{T}{M^2})  
\label{w1} 
\eeq 
where $g$ is any function of $T$ but not $\bar{T}$.  
When we impose both nonanomalous combinations  
of symmetries, we constrain the form of the superpotential further to the  
form  
\beq 
W= M^5 \frac{e^{-2 \pi S} }{T}  
\label{w2} 
\eeq 
which is exactly the form of the nonperturbative  
superpotential for $SU(2)$ which we started the discussion with.  
 
It should be observed, that the crucial role in determining the  
S-dependence of the superpotential is played by the imaginary shift in S.  
In the above, we have assumed that that shift should be continuous,  
but experience with strings tells us, that it should rather be discreet. 
With our present normalization of S the discrete shifts allowed by  
the string are  
\beq 
S \rightarrow S - i n 
\label{ds} 
\eeq 
where $n$ is integer.  
Then the nonanomalous remnant of $U(1)_A$ is  
\beq 
S\rightarrow S-i n,\; Q \rightarrow e^{-i q \frac{n}{\ks}} Q,\;  
\bar{Q} \rightarrow e^{-i \bar{q} \frac{n}{\ks}} \bar{Q} 
\label{dnac} 
\eeq 
This discreete symmetry is much less restrictive and  
allows the general superpotential  
\beq 
W = f(e^{-2 \pi S}, T)  
\label{dre} 
\eeq 
where f is any function restricted by analycity requirements  
and by the requirement that the superpotential falls down 
exponentially in the weak coupling limit (one  expects
 nonperturbative contributions to  
be proportional to expenentials with negative exponents - like $-8 \pi/g^2$ 
in one-instanton case). 
It is easy to see that the discrete version of $U(1)_{R''}$ doesn't  
restrict further the S-dependence of the superpotential. 
However, if we impose $U(1)_{R'}$, then we get  
\beq 
W= \frac{h(e^{-2 \pi S})}{T} 
\eeq 
which is  
 the most general form allowed.  
Hence, for instance, string allows the effective superpotential  
to be a series $ \sum_{i} c_{i} e^{- k_i 2 \pi \frac{S}{M}}$ where  
$k_i$ are integers, which can lead to stabilization of the dilaton 
 (from here on we are back to the dimensionful version of  
the superfield S).  
 
Further to that, the discreete imaginary shift  
can be viewed as the part of the  
larger symmetry group, for instance as a subgroup of  
$SL(2,Z)$ S-duality symmetry  
as discussed in \cite{b1}. This leads to the construction of more  
specific superpotentials, which can be seen as an infinite series of  
exponentials. An example of the S-dual model is 
\beq 
W(S,T) = \frac{\alpha}{T \eta^2 ( \frac{S}{M})  
(j(\frac{S}{M})-744)^{|\beta|}}  
\label{sd1} 
\eeq 
where $\eta$ is the modular $\eta$-form, $j$ denotes the modular invariant  
$j$-function, and $\alpha, \; \beta$ are constants, 
with the gauge coupling given by   
\beq 
g^2 = \frac{8 \pi^2}{\log |j(\frac{S}{M})-744|} 
\eeq  
Model given with these functions in the absence of the anomalous $U(1)$ 
would have a minimum in $S$ at the self-dual point $S=1$\footnote{Of  
course, 
to fix $T$ we need also here a $T$-dependent perturbation of the $S$-dual  
superpotential}, with supersymmetry unbroken in the direction of S.   
 
\section{Exponential superpotentials} 
 
To have some idea of what the actual orders of magnitude of the parameters  
in the realistic ``binding'' superpotentials can be, let us consider  
more detaily the example of 
superpotentials which can be considered as series of exponential terms, 
in agreement with the discussion of the previous section 
\beq 
W(S) = \sum_{\gamma} c_{\gamma} e^{- \gamma 2 \pi \frac{S}{M}} 
\eeq 
where the sum is finite, as in race-track models, or infinite,  
as in S-dual models of \cite{b1}. 
In any case we shall assume that there is a supersymmetric point  
$S_o$ in these models, which corresponds to a minimum in the  
globally supersymmetric model with the assumed superpotential.  
The choice of the $S_o$ to be the supersymmetric point is dictated  
by the experience with string-inspired models. In these models  
at the natural minima of the dilaton supersymmetry is typically unbroken, 
or only slightly broken, along the dilaton direction.  
The general expansion is  
\beq 
W(S)= W(S_o) + \frac{\partial W}{\partial S} (S - S_o) +  
\frac{\partial^2 W}{\partial S^2} (S - S_o)^2 + ... 
\label{gauge} 
\eeq 
where the second term on the $rhs$ vanishes by assumption.  
When we apply this expansion to exponential superpotentials,  
and truncate it after the second term, 
we get  
\beq 
W(S)= \Lambda^3  + c \frac{\Lambda^3}{M^2} (S - S_o)^2  
\eeq 
In the above formula the constant $c$ is of order unity in race-track 
models, but can be larger, eg. $O(10)$ or higher, in S-dual models. 
Of course, the scale of $\Lambda$ sets the gravitino mass, $m_{3/2}  
\approx \frac{\Lambda^3}{M^2}$, and $c \frac{\Lambda^3}{M^2}$ is  
the scale of the supersymmetric contribution to the dilaton mass.  
 
Hence, in the notation of the previous chapter we obtain 
\beqa 
q&=& c \frac{\Lambda^3}{M^2} \nonumber \\ 
p&=& S_o 
\eeqa 
Now, it is easy to find out the realistic values of $m$ 
looking at the generalized version of (\ref{pert}) 
\beq 
W_{pert} = \frac{(G_{-} G_{+})^n X_{-} X_{+}}{M^{2 n -1}} \lambda 
\label{gpert} 
\eeq 
In this case the perturbative $U(1)_A$ invariance of the superpotential  
requires that $q_{+}+q_{-}= -(q+\bar{q}) n = 2 \pi \ks n$ 
which means that we should replace in all relevant formulae $q_{+} \rightarrow 
2 \pi \ks n +1$. 
Now we can identfy $m$ as 
\beq 
m= \frac{<G_{-} G_{+}>^n }{M^{2 n -1}} \equiv  
\frac{<T>^n }{M^{2 n -1}}  
\eeq 
Now, as we have mentioned before, one can probably achieve  
$<T>=M^2$ in supergravity models. In such case $m=M \gg q$. 
However, the natural value of $T$, consistent with the observation  
that formation of condensates due to strong gauge forces is rather a 
field theoretical phenomenon, is $<T>=\Lambda^2$.  
In this case, if n=1 we obtain $m= \Lambda \frac{\Lambda}{M} = 
\frac{M}{\Lambda} q \gg q$.  
For all $n > 1$ the hierarchy between $q$ and $m$ gets inverted - 
$m$ becomes much smaller than $q$.  
It is a straightforward excercise to obtain values of F- and D-terms  
in all the cases.  
 
\begin{table}
\begin{center} 
\begin{tabular}{||c|c|c|c||}  
\hline \hline 
{} & $n=1$ & $n=2$ & $n >  2$  
\\ \hline \hline 
$m$ & $\Lambda \frac{\Lambda}{M}$ &$\Lambda (\frac{\Lambda}{M})^3$&  
$\Lambda (\frac{\Lambda}{M})^{2n -1}$ \\ \hline 
$q$ & \multicolumn{3}{c||}{$c \Lambda (\frac{\Lambda}{M})^2$} \\ \hline 
$\frac{m}{q}$&$\frac{1}{c} (\frac{M}{\Lambda}) \gg 1$& $\frac{1}{c} 
 (\frac{\Lambda}{M}) \ll 1$& $\frac{1}{c} 
 (\frac{\Lambda}{M})^{2 n -3} \ll 1$\\ \hline  
$\frac{|F_{\x}|}{|F_{\y}|}$& $ \frac{<s>}{ 2 \pi M}$& 
 \multicolumn{2}{c||}{$\frac{m}{q} \ll 1$}\\ \hline 
$m_{3/2}^2 = M^2 e^{-G}$&  \multicolumn{3}{c||}{$\Lambda^2  
(\frac{\Lambda}{M})^4$} \\ \hline 
$\frac{\delta_f m^2}{\delta_d m^2}$&$\ks \frac{<s>}{\pi M}$&  
\multicolumn{2}{c||}{$\ks \frac{M}{p}$}\\ \hline 
$m^2_{soft}$& $\ks \frac{<s>}{\pi M} \Lambda^2 (\frac{\Lambda}{M})^2 
 >m_{3/2}^2 $&$ \Lambda^2 (\frac{\Lambda}{M})^6$&$  
\Lambda^2 (\frac{\Lambda}{M})^{4 n -2}$ \\ \hline \hline  
\end{tabular} 
\end{center} 
\caption{Summary of the models with typical exponential superpotentials. 
The values of $c \approx 1$ and of $p \leq 1$ in the column corresponding to  
$n=1$ are assumed.} 
\label{tb:fit1} 
\end{table} 
 
The Table (\ref{tb:fit1}) summarizes the models with generic  
exponential superpotentials. It is obvious that the most interesting models  
correspond to the $n=1$. In this case, for $p \leq 20$ and $\Lambda /M \leq 
10^{-5}$ - which is just the interesting range of parameters,  
the results closely follow those for the generic model II.  
In fact, to a very good approximation the minimum of the effective  
potential resulting from integrating out all the degrees of freedom  
except $S$ is 
\beq 
<S_o> = M (\frac{\ks \pi}{16})^{1/5} (\frac{M}{\Lambda})^{2/5} 
\eeq 
If one demands that the assumed minimum corresponds to the value  
given by the unification conditions, then one obtains a simple  
 relation  
between the trace of the anomalous $U(1)_A$ and the condensation scale  
in the hidden sector   
\beq 
\log (|Tr X|) = \; {const}\; + \log(\frac{\Lambda}{M}) 
\label{fin} 
\eeq 
It is apparent that for a given strongly interacting hidden sector  
one can work on the charges of the $U(1)$ group towards achieving  
required value of the unification coupling. Once this is done, 
one obtains highly universal, in the sense discussed in the introduction,  
 soft masses which are generated in  
similar parts by the non-zero D-term and by the dilaton F-term. In this  
scenario also the gaugino masses are large and equal in magnitude to  
scalar masses.   

\section{Inflation from dynamical F-I term}

Some cautious remarks must be made about cosmology of the models we  
discuss here. As the global supersymmetry is broken in these models 
 spontaneously  
in the flat limit, 
they generically have a positive, and unacceptably large, cosmological  
constant. Cancellation of this constant through the gravitational corrections, 
if possible at all, would require for instance an addition of the constant 
 piece in the superpotential of the order $ <F> M$ at least, which  
is hardly natural in the present context. The dynamical dilaton in  
the D-term implies also harmful modifications to  scenarios of  
 inflation based on the temporary dominance of the D-term in the potential 
energy \cite{b12}. 
To illustrate the trouble let us consider the epoch when the D-term dominates 
the energy density. The evolution equations for the scale factor $a(t)$ 
and the dilaton $s(t)$ are (we put $M=1$ here)
\beqa
\frac{\partial^2 s}{\partial t^2} - (\frac{\partial s}{\partial t})^2 /s 
+ 3 H \frac{\partial s}{\partial t} - \frac{ 192 \pi^3 \delta_{GS}^2}{s^2}&=& 0
\nonumber \\
(\frac{\partial s}{\partial t})^2 /( 4 s^2)  + \frac{32 \pi^2 \delta_{GS}^2}
{s^3}& = & H^2
\eeqa
where $H$ is the Hubble parameter.  
The results of the integration of these equations for $\delta_{GS} = 0.1$ 
is shown in the 
figure 1. 
The figure shows that there is no inflationary epoch after a short 
initial switching-on period which is due to the initial conditions taken 
for the integration. The dilaton evolves so rapidly, that its kinetic 
energy very quickly comes to dominate over the potential energy - 
which doesn't allow   
inflation. The large kinetic energy which dilaton would acquire during 
the stage of such a D-term dominated evolution is very likely to 
prevent it from settling down in any conceivable minimum which could be generated 
by gauge dynamics in the intermediate or weak coupling regime.      

\par
\centerline{\hbox{
\psfig{figure=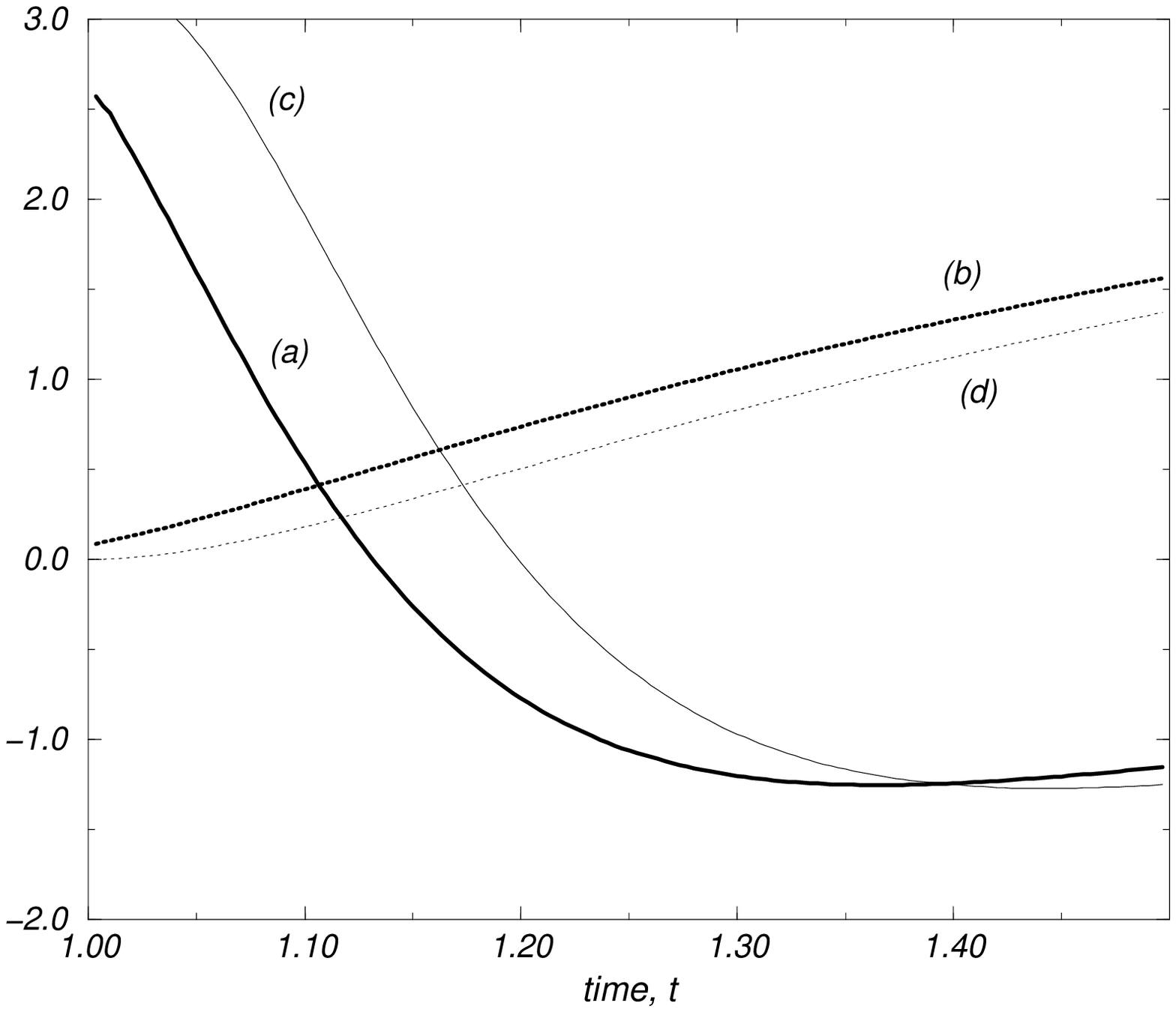,height=4.0in,width=5.0in}
}}
\par
{\small \em Figure 1. Evolution of the acceleration of 
the scale factor, curves 
(a),(c), and of the ratio of the dilaton kinetic energy and the D-term energy,
curves (b) and (d), in the Universe dominated initially by the planck scale 
dynamical D-term. Thin curves correspond to the solution with the vanishing initial velocity of the dilaton. and the thick curves to the planck scale initial velocity
of the dilaton. In both cases the acceleration becomes quickly negative, and 
the kinetic energy of the dilaton comes to dominate expansion.}\\

\noindent{ }
 
\section{Discussion and conclusions} 
 
In this paper we have analyzed in as much model independent manner as  
possible the structure of the dilaton vacuum and supersymmetry breaking  
pattern in Fayet-Iliopoulos models with dynamical gauge coupling  
and large (in fact - stringy) value of the F-I parameter. 
 
We have found that in this important class of models the supersymmetry  
is generically broken, but the mere 
presence of the D-term is not sufficient to 
stop the running away of the modulus responsible for the value of the  
gauge coupling - the dilaton.  

Some of the models described in section 4 have an interesting vacuum structure resembling softly broken SQCD. They can serve as a  secluded sector 
in models with gauge mediation of the susy breaking once the dilaton gets stabilized. 
 
To stabilize the dilaton, one has to include an additional dilaton-dependent  
part in the superpotential. Once this is the case, several scenarios are  
possible, as described in the preceding section.  
In particular, the models which without the D-term would be  
considered irrelevant as giving strongly coupled gauge theories  
at the unification scale, can be saved in the presence of the  
large D-term. Such models, when asked to give the correct  
$g^2_{unification}$, easily give rise to the mixed dilaton/D-term  
dominated scenarios of susy breaking, which in principle allow  
horizontal hierarchy generation. The models with purely dilaton 
dominated supersymmetry breaking could also be thought of, but they  
would correspond to very weak unified coupling (cf. Table (\ref{tb:fit1}), 
column $n=1$).  
  
It should be noted, that in general one cannot assume $F^{\x}=0$  
in the class of models discussed here. 
 
Finally, as easily seen from the Table (\ref{tb:fit1}) that all interesting effects  
of the d-term are proportional to the value of $\ks$ 
 hence quickly become unimportant when the mass scale of the F-I 
parameter becomes smaller than $M_{planck}$. Also, after supersymmetry 
breaking 
induced by some third party one can have induced d-terms with F-I effective  
parameters $\xi \sim m^2_{3/2} \log m^2_{3/2}$ which for reasonable  
values of $m^2_{3/2} \sim 1 TeV^2$ gives $\frac{\xi}{M^2} \leq 10^{-30}$  
resizing all the interesting features discussed above down to nothing.  
 
Some cautious remarks must be made about cosmology of the models we  
discuss here. As the global supersymmetry is broken in these models 
 spontaneously  
in the flat limit, 
they generically have a positive, and unacceptably large, cosmological  
constant. Cancellation of this constant through the gravitational corrections, 
if possible at all, would require for instance an addition of the constant 
 piece in the superpotential of the order $ <F> M$ at least, which  
is hardly natural in the present context. 
We have shown that when the F-I parameter and the gauge coupling are dynamical 
variables, the D-term dominated Universe does not allow for an inflationary period.\\
 
\noindent{\bf Acknowledgments} 
This work has been supported in part
by Polish Commitee for Scientific Research grant 2 P03B 040 12, and 
by M.Curie-Sklodowska Foundation and Polish-French Collaborative Projects 
Programme. Z.L. is also grateful for the hospitality  
of the Center of Theoretical Physics of Seoul National University where part of the work was done. \\
Z.L. would like to thank M. Spali\'nski for discussions and collaboration 
at the early stage of the project.\\ 
This work was presented at the Third Warsaw Workshop ``Physics from Planck Scale to Electroweak Scale'', Warsaw, 2-5 April 1997.\\ 
\noindent{\bf Added note} We got aware of the eprint \cite{b14}
which also adresses some of the issues considered here.\\

\end{document}